\documentstyle[prd,eqsecnum,aps,epsf]{revtex}

\tightenlines

\newcommand{\be}{\begin{eqnarray}}
\newcommand{\beq}{\begin{eqnarray}}
\newcommand{\befg}{\begin{figure}}
\newcommand{\CDP}{\hat{\cal D}\hspace{-0.27cm}\slash_v}
\newcommand{\CDPL}{\overleftarrow{\hat{\cal D}}\hspace{-0.29cm}\slash_v}
\newcommand{\CDPN}{{\cal D}\hspace{-0.26cm}\slash_v}
\newcommand{\CDPNL}{\overleftarrow{\cal D}\hspace{-0.29cm}\slash_v}

\newcommand{\DS}{D\hspace{-0.25cm}\slash}
\newcommand{\DSP}{\not\!{v} v\cdot D}
\newcommand{\dsp}{D\hspace{-0.27cm}\slash_{\|}}
\newcommand{\DSC}{D\hspace{-0.26cm}\slash_{\bot}}
\newcommand{\DSCX}{D\hspace{-0.26cm}\slash_{\bot}}
\newcommand{\DSCXL}{\overleftarrow{D}\hspace{-0.29cm}\slash_{\bot}}

\newcommand{\edfg}{\end{figure}}
\newcommand{\eeq}{\end{eqnarray}}
\newcommand{\ee}{\end{eqnarray}}

\newcommand{\MQ}{m_Q}

\newcommand{\oDSP}{\not\!{v} v\cdot \overleftarrow{D}}
\newcommand{\odsp}{\overleftarrow{D}\hspace{-0.29cm}\slash_{\|}}
\newcommand{\oDslashbot}{\overleftarrow{/\!\!\!\!D}\hspace{-5pt}_{\bot}}

\newcommand{\PP}{{1 + v\hspace{-0.2cm}\slash \over 2}}
\newcommand{\PM}{{1 - v\hspace{-0.2cm}\slash \over 2}}
\newcommand{\PSC}{\partial\hspace{-0.2cm}\slash_{\bot}}
\newcommand{\PSP}{\partial\hspace{-0.2cm}\slash_{\|}}

\newcommand{\qpv}{Q_v^{(+)}}
\newcommand{\qpvb}{\bar{Q}_v^{(+)}}

\newcommand{\QVH}{\hat{Q}_v}
\newcommand{\QVHB}{\bar{\hat{Q}}_v}
\newcommand{\QVHF}{\hat{Q}^{(-)}_v}
\newcommand{\QVHMP}{\hat{Q}^{(\mp)}_v}
\newcommand{\QVHFB}{\bar{\hat{Q}}_v{\vspace{-0.3cm}\hspace{-0.2cm}{^{(-)}} }}
\newcommand{\QVHPM}{\hat{Q}^{(\pm)}_v}

\newcommand{\QVHPMB}{\bar{\hat{Q}}_v{\vspace{-0.3cm}\hspace{-0.2cm}{^{(\pm)}} }}
\newcommand{\QVHZB}{\bar{\hat{Q}}_v{\vspace{-0.3cm}\hspace{-0.2cm}{^{(+)}} } }
\newcommand{\QVHZ}{\hat{Q}^{(+)}_v}

\newcommand{\uvslash}{/\!\!\!\!\hspace{1pt}v}
\newcommand{\VS}{\not\!{v}}

\begin{document}
\title{HQEFT as A Large Component QCD and \\ Comments on The Incompleteness of HQET}
\author{ Y.L. Wu, Y.A. Yan, M. Zhong and Y.B. Zuo }
\address{ Institute of Theoretical Physics, Chinese Academy of Sciences, \\
P.O. Box 2735, Beijing 100080, China }
\author{W. Y. Wang}
 \address{ Department of Physics, Tsinghua University, Beijing 100084, China }
\maketitle
\begin{abstract}
The Heavy quark effective field theory (HQEFT) is revisited in a
more intuitive way. It is shown that HQEFT is a
consistent large component QCD of heavy quarks. In the non-relativistic limit,
HQEFT recovers the non-relativistic QCD (NRQCD). The resulting new
effects in the HQEFT of QCD are carefully reexamined. It is then
natural to come to the comments on the usual heavy quark effective
theory (HQET). Consistent phenomenological
applications of HQEFT exhibit its interesting features and
completeness in comparison with HQET. It then becomes manifest
why we shall base on the HQEFT of QCD rather than HQET which is
an incomplete one for computing $1/m_Q$ corrections. More precise extraction
for $|V_{cb}|$ and $|V_{ub}|$ in the HQEFT of QCD is emphasized.
\end{abstract}



\section{Introduction}

  Two frameworks of effective theories for heavy quarks have extensively been applied to treat
  heavy quark systems containing a single heavy quark.
  One is so-called heavy quark effective theory (HQET)\cite{HG,HQ} and the other
  is mentioned, for distinction, to be the heavy quark effective field
  theory (HQEFT)\cite{YLW}. The former was motivated and constructed mainly
  based on the heavy quark symmetry (HQS)\cite{symmetry,symmetry2,Is.Wi}
  in the infinity mass limit $m_Q\rightarrow \infty$\cite{VS,EH}.
  The latter was derived directly from QCD with carefully treating
  the quark and anti-quark fields.
 In the HQET, the quark and antiquark fields were treated separately,
 which is valid only in the infinity mass limit. For the finite quark mass case,
 it is well known from the quantum field theory that the quark and antiquark fields
 cannot be separated though their interactions may be suppressed by the inverse of the mass.
 This is one of the main reasons that motivated one of our authors to consider and
 derive an alternative HQEFT from QCD with explicitly keeping the quark and antiquark
 coupling interactions\cite{YLW}. In deriving the HQEFT,
 the approximate spin-flavor symmetry is not as the starting point but
 as a direct consequence in the heavy quark
 limit. In fact, it is the approximate
 spin-flavor symmetry that has led a rapid development on
 effective theories of heavy quark\cite{2}-\cite{12}
 in studying heavy quark systems.
 It is obvious that the two frameworks, HQET and HQEFT of QCD, become equivalent
 only when taking the heavy quark mass to be infinite. This is because that
only in this case the quark and antiquark can be completely decoupled.
 The possible differences between them have recently been investigated
 in detail from heavy hadron decays\cite{W1}-\cite{W10}, where
 the important new effects have been found to
 arise from the $1/m_Q$ corrections. The interesting applications of HQEFT have also been studied in
 the pair creations and annihilations\cite{bs1,bs2,mas}. In this paper, we are going to revisit the HQEFT
 of QCD in a more intuitive way and make it more clear for acceptance. Especially, one will
 see why HQEFT
 should be regarded as a more complete effective theory of QCD for heavy quark systems.
 To be manifest, we consider two interesting cases that can be dealt with by the HQEFT of QCD:
 one is for a heavy quark within the heavy quarkonia systems, which is known to be treated
 by the non-relativistic QCD (NRQCD)\cite{wg,bbl}; the other is for a single heavy quark
 within the heavy hadron systems\cite{W1}-\cite{W10}. Some comments on
 the HQET then become obvious. Several important and interesting
 observations yielded in the HQEFT of QCD
 are briefly summarized. The emphasis is made to the more precise extractions on the
 Cabbibo-Kobayashi-Maskawa (CKM) matrix elements $V_{cb}$ and $V_{ub}$ via HQEFT of QCD.

\section{Why HQEFT of QCD rather than HQET}

HQEFT of QCD is a theoretical framework derived directly from QCD
but explicitly displays the heavy quark symmetry and symmetry
breaking corrections. The main point is that all contributions of
the field components, large and small, `particle' and
`antiparticle', have carefully been treated in the Lagrangian of
HQEFT, so that the resulting effective Lagrangian form the basis
for a ``complete" effective field theory of heavy quarks
\cite{YLW,W1,W2,W3,W4,W5,W6,W7,W8,W9,W10}. We will see that HQEFT
of QCD is actually an effective field theory of large component
QCD.

 Let us begin with the Lagrangian of full QCD
\begin{equation}
\label{QCDL}
{\cal L}_{QCD}={\cal L}_{light} + \bar{Q} (i\DS-m_Q) Q,
\end{equation}
where $Q$ is the quantum field for heavy quark and ${\cal
L}_{light}$ represents the part that has nothing to do with the
heavy quark. Following the derivation presented in ref.\cite{YLW},
the field $Q$ can formally be written into two part
\begin{equation}
\label{eq:7}
    Q=Q^{(+)}+Q^{(-)} ,
\end{equation}
where $Q^{(+)}$ and $Q^{(-)}$ are classically corresponding to two
solutions of Dirac equation
     \begin{equation}
     \label{eqmotion}
         (iD\hspace{-0.25cm}\slash-\MQ)Q^{(\pm)}=0 .
     \end{equation}
For the free quark fields, they are so called ``quark" and
``antiquark" fields respectively, and can be expanded in terms of
plane waves
\begin{eqnarray}
Q^{(+)}(x) &=& \int \frac{d^3p}{(2\pi)^3}{m\over p^0}
    \sum\limits_s b_s(p)u_s(p)e^{-ip\cdot x},  \\
Q^{(-)}(x) &=& \int \frac{d^3p}{(2\pi)^3}{m\over p^0}
    \sum\limits_s d^{\dag}_s(p)v_s(p)e^{ip\cdot x},
\end{eqnarray}
where $s$ is spin index, $b_s$ and $d_s$ are the annihilation and
creation operators respectively. $u_s$ and $v_s$ are
four-component spinors. In Dirac representation, they can be
explicitly written as
\begin{eqnarray}
u_s(p) &=& \sqrt{E + m \over 2m}
\left( \begin{array}{c} 1 \\ {{\bf \sigma \cdot p}\over E+m} \end{array}
\right)
\varphi_s ,  \\
v_s(p) &=& \sqrt{E + m \over 2m}
\left( \begin{array}{c} {{\bf \sigma \cdot p}\over E+m} \\ 1 \end{array}
\right)
\chi_s
\end{eqnarray}
with $\varphi_s$ being the two component Pauli spinor field that annihilates a heavy quark,
and $\chi_s$ being the Pauli spinor field that creates a heavy antiquark.

Introducing an arbitrary unit vector $v^\mu$ which satisfies $v^2
= 1$, we can further decompose the quark fields $Q^{(\pm)}$ into
the following forms
\begin{eqnarray}
\label{Qdec1}
Q^{(+)} &\equiv& \Big({1 + \uvslash \over 2} + {1 - \uvslash \over 2} \Big) Q^{(+)}
= \hat{Q}^{(+)}_v + R^{(+)}_v,  \\
\label{Qdec2}
Q^{(-)} &\equiv& \Big({1 - \uvslash \over 2} + {1 + \uvslash \over 2}\Big) Q^{(-)}
= \hat{Q}^{(-)}_v + R^{(-)}_v
\end{eqnarray}
with
    \begin{equation}
    \label{eq:9}
      \hat{Q}^{(\pm)}_v\equiv \frac{1{\pm}v\hspace{-0.2cm}\slash}{2}Q^{(\pm)},\hspace{1.5cm}
      R^{(\pm)}_v\equiv \frac{1{\mp}v\hspace{-0.2cm}\slash}{2}Q^{(\pm)} .
    \end{equation}
 Thus the initial quark field $Q$ can be written as
 \begin{equation}
 Q \equiv \hat{Q}_v + R_v
 \end{equation}
 with
 \begin{equation}
 \hat{Q}_v = \hat{Q}_v^{(+)} + \hat{Q}_v^{(-)},
 \qquad R_v = R_v^{(+)} + R_v^{(-)} .
 \end{equation}
For free quark field, taking $v=(1, 0, 0, 0)$, one has in the momentum space
\begin{eqnarray}
\label{component1}
\hat{Q}_v^{(+)} \to \PP u_s(p) &=& \sqrt{E + m \over 2m}
\left( \begin{array}{c} 1 \\ 0 \end{array} \right)
\varphi_s ,
\\
\label{component2}
R_v^{(+)} \to \PM u_s(p) &=& \sqrt{E + m \over 2m}
\left( \begin{array}{c} 0 \\ {{\bf \sigma \cdot p}\over E+m} \end{array}
\right)
\varphi_s ,
\\
\label{component3}
R_v^{(-)} \to \PP v_s(p) &=& \sqrt{E + m \over 2m}
\left( \begin{array}{c}{{\bf \sigma \cdot p}\over E+m} \\ 0 \end{array} \right)
\chi_s ,
\\
\label{component4}
\hat{Q}^{(-)}_v \to \PM v_s(p) &=& \sqrt{E + m \over 2m}
\left( \begin{array}{c} 0 \\ 1 \end{array} \right)
\chi_s .
\end{eqnarray}
It is then obvious from Eqs.(\ref{component1})-(\ref{component4}) that $\hat{Q}_v^{(+)}$ and $\hat{Q}_v^{(-)}$ may
be regarded as the ``large components" of ``quark" and ``antiquark" respectively, while $R_v^{(+)}$ and
$R_v^{(-)}$ are the corresponding ``small components" in the case $ |{\bf p}| \ll E + m $.

Here it should be noted that though the ``large" and ``small"
components of quark and antiquark fields can be understood easily
in this way by taking a special vector $v=(1,0,0,0)$, the vector
$v$ does not have to be chosen in this way. In fact, rewriting the
derivative operator as
\begin{eqnarray}
    \label{eq:4}
 \DS = \dsp +  \DSC
\end{eqnarray}
with
\begin{equation}
 \dsp \equiv \DSP, \quad \DSC \equiv \DS-\VS (v\cdot D) ,
\end{equation}
 and using Eq.(\ref{eqmotion}) as well as the commutation relations
\begin{eqnarray}
 \label{commutation}
[\VS, \dsp]=0, \qquad \{\VS,  \DSC \}=0,
\end{eqnarray}
 one can easily find the relations between the fields $R^{(\pm)}_v$ ($\bar{R}^{(\pm)}_v$) and $\QVHPM$
($\QVHPMB$) :
\begin{eqnarray}
\label{RinQVH}
(i\dsp - m_Q) R_v^{(\pm)}  + i\DSC \hat{Q}_v^{(\pm)} = 0 , \\
\bar{R}^{(\pm)}_v (-i\odsp - m_Q)  - \QVHPMB i\oDslashbot = 0 .
\end{eqnarray}
For any operator $O$, the operator $\overleftarrow{O}$ is defined via
 $\int\kappa\overleftarrow{O}\varphi \equiv -\int\kappa O \varphi$.
Thus the fields $Q$ and $\bar{Q}$ can be represented by $\hat{Q}_v$ and $\bar{\hat{Q}}_v$,
     \begin{eqnarray}
     \label{QinQVH1}
         Q=\Big[1+\Big(1-\frac{i\DSP+\MQ}{2\MQ}\Big)^{-1}
           \frac{i\DSCX}{2\MQ}\Big] \hat{Q}_{v}\equiv\hat{\omega} \hat{Q}_v ,\\
 \label{QinQVH2}
         \bar{Q} = \bar{\hat{Q}}_{v}\Big[1+
        \frac{-i
\DSCXL
}{2\MQ}\Big(1-\frac{-i\oDSP+\MQ}{2\MQ}\Big)^{-1}\Big]
\equiv \bar{\hat{Q}}_v\overleftarrow
           {\hat{\omega}}.
     \end{eqnarray}
From Eqs.(\ref{Qdec1}), (\ref{Qdec2}) and (\ref{QinQVH1}), (\ref{QinQVH2}) one can see that, in this
representation, $R_v$ is $1/m_Q$ suppressed compared to $\QVH$. Therefore $\QVHPM$ and $R^{(\pm)}_v$ can be
regarded as the ``large component" and ``small component" of the heavy quark field $Q^{(\pm)}$ respectively. This is
independent of the special choice of vector $v^{\mu}$ as long as it satisfies the condition $v^2=1$.

 To construct the effective field theory of large component $\hat{Q}_v$, one can simply integrate out
 the small component $R_v$, or integrate in the contributions of small component to the effective
 Lagrangian. This has been shown\cite{W1,W2} to be
 equivalent to substitute Eqs.(\ref{QinQVH1}),(\ref{QinQVH2}) into the QCD Lagrangian. Thus the resulting large component
 QCD can be written as
    \begin{equation}
    \label{eq:13}
       {\cal L}_{QCD} ={\cal L}_{light}+{\cal L}_{Q,v}
    \end{equation}
with
    \begin{eqnarray}
    \label{eq:14}
      {\cal L}_{Q,v}& \equiv&\bar{Q}(i\DS-\MQ)Q |_{Q\to \hat{\omega}\hat{Q}_v} \nonumber\\
   & =& \QVHB(i\DSP-\MQ)\QVH+\frac{1}{2\MQ}\QVHB i\DSC
      \Big(1-\frac{i\DSP+\MQ}{2\MQ}\Big)^{-1}i\DSC \QVH \nonumber \\
   & + & \frac{1}{2\MQ} \QVHB i\DSC \Big(1-\frac{i\DSP+\MQ}{2\MQ} \Big)^{-1}
      (i\DSP-\MQ) \QVH  \\
   & + & \frac{1}{4\MQ^2}\QVHB (-i
      \DSCXL) \Big(1-\frac{-i\oDSP
      +\MQ}{2\MQ}\Big)^{-1}i\DSC \Big(1-\frac{i\DSP+\MQ}{2\MQ}\Big)^{-1}i\DSC \QVH . \nonumber
    \end{eqnarray}

From the definitions (\ref{eq:9}) and the commutation relations (\ref{commutation}),
one can see that there is no quark-antiquark coupling in the first two terms on the rhs. of Eq.(\ref{eq:14}). The
last two terms on the rhs. of Eq.(\ref{eq:14}) arise from the quark-antiquark coupling interactions, which has
been ignored in the HQET Lagrangian. One can rewrite the above Lagrangian ${\cal L}_{Q,v}$ in the following form
to see explicitly such features,
   \begin{equation}
     \label{lagr}
        {\cal L}_{Q,v} = {\cal L}^{(++)}_{Q,v}+{\cal L}^{(--)}_{Q,v}
          +{\cal L}^{(+-)}_{Q,v}+{\cal L}^{(-+)}_{Q,v}
     \end{equation}
with
    \begin{eqnarray}
    \label{lzz}
{\cal L}^{(\pm \pm)}_{Q,v} &=&
        \QVHPMB [i\CDP -\MQ] \QVHPM , \\
 \label{lzf}
{\cal L}^{(\pm \mp)}_{Q,v} & = &
              \frac{1}{2\MQ} \QVHPMB (i \CDPL -\MQ) \Big(1-\frac{i\DSP+\MQ}{2\MQ}\Big)^{-1} (i\DSCX )\QVHMP \nonumber \\
        & = & \frac{1}{2\MQ} \QVHPMB (-i\DSCXL)\Big(1-\frac{-i\oDSP+\MQ}{2\MQ}
   \Big)^{-1}(i\CDP -\MQ) \QVHMP,
\end{eqnarray}
where the operator $i\CDP$ is defined as
\begin{eqnarray}
i\CDP &=& i\DSP + \frac{1}{2\MQ}i\DSC \Big(1-\frac{i\DSP+\MQ}{2\MQ} \Big)^{-1} i\DSC .
\end{eqnarray}
The operator $\CDPL$ in the above equations can be obtained by
replacing $D^{\mu}$ with $-\overleftarrow{D^{\mu}}$ in $\CDP$.
Note that the above Lagrangian actually holds for either heavy or
light quarks as it is obtained by just integrating out the field
 component $R_v$. For heavy quarks with $ |{\bf p}| \ll E + \MQ$, one can
then make expansion in terms of inverse powers of heavy quark mass
$\MQ$. The resulting effective theory is called HQEFT. It is
different from the HQET in which only ${\cal L}^{(++)}_{Q,v}$ (or
${\cal L}^{(--)}_{Q,v})$ in Eq.(\ref{lagr}) is taken to be the
effective Lagrangian. Obviously, the HQET is not complete. In
fact, in any physical process of heavy quark, there is heavy
quark-antiquark coupling interactions due to the finite mass of
heavy quark. Such quark-antiquark coupling terms will provide
 important contributions at the order of $1/m_Q$. Therefore in order to calculate
correctly these $1/m_Q$ corrections arising from the finite mass
of heavy quark, one should take into account the contributions
from the terms ${\cal L}^{(\pm\mp)}_{Q,v}$ and ${\cal
L}^{(--)}_{Q,v}$ for the processes involving the heavy quark
$Q^{(+)}_v$ or from the terms ${\cal L}^{(\pm\mp)}_{Q,v}$ and
${\cal L}^{(++)}_{Q,v}$ for the ones involving the heavy antiquark
$Q^{(-)}_v$ .

  For making $1/m_Q$ expansion, it is useful to remove the large mass term in the above Lagrangian. We then introduce
new field variables $Q_v$ and $\bar{Q}_v$ with the definition
    \begin{equation}
    \label{eq:21}
       Q_v=e^{iv\hspace{-0.15cm}\slash m_Q v\cdot x}\QVH ,\hspace{1.5cm}
       \bar{Q}_v=\QVHB e^{-i v\hspace{-0.15cm}\slash m_Q v \cdot x} .
    \end{equation}
Noticing the feature that $\VS$ commutes with $\dsp$ but
anticommutes with $\DSC$,  we can rewrite above Lagrangian to be
\begin{eqnarray}
    \label{lzz2}
{\cal L}^{(\pm \pm)}_{Q,v} &=&
        \bar{Q}_v i\CDPN Q_v , \\
 \label{lzf2}
{\cal L}^{(\pm \mp)}_{Q,v} & = &
              \frac{1}{2\MQ} \bar{Q}_v (i\CDPNL)  e^{2iv\hspace{-0.15cm}\slash m_Q v\cdot x}
        \Big(1-\frac{i\DSP}{2\MQ}\Big)^{-1} (i\DSCX ) Q_v \nonumber \\
        & = & \frac{1}{2\MQ} \bar{Q}_v (-i\DSCXL) \Big(1-\frac{-i\oDSP}{2\MQ}
    \Big)^{-1} e^{-2iv\hspace{-0.15cm}\slash m_Q v\cdot x}(i\CDPN) Q_v
\end{eqnarray}
with
\begin{eqnarray}
& & i\CDPN=i\DSP+\frac{1}{2\MQ}i\DSC \Big(1-\frac{i\DSP}{2\MQ} \Big)^{-1} i\DSC ,
   \nonumber \\
 & & i\CDPNL = -i\oDSP +
 \frac{1}{2\MQ}   (-i\DSCXL)\Big(1-\frac{-i\oDSP}{2\MQ}\Big)^{-1} (-i\DSCXL) ,
\end{eqnarray}
 where we have expressed the effective Lagrangian for quark-antiquark coupling interactions
 (\ref{lzf2}) in terms of two identical formulations. One can use
 either of them depending on the convenience for the relevant
 applications.

 The above Lagrangian (eqs.(2.26-29) or eqs.(2.31-33)) forms the basic framework of
 HQEFT which has first been derived in ref.\cite{YLW}. Here we rewrite it in a more compact expression.
 Note that the resulting effective Lagrangian is exact for the quark component
 $Q_v=Q_v^{(+)} + Q_v^{(-)}$. What we have done is making the
 field redefinitions and integrating out the quark field component $R_v=R_v^{(+)} + R_v^{(-)}$, or
 integrating in the contributions of the quark component $R_v=R_v^{(+)} + R_v^{(-)}$
 to the effective Lagrangian.
 It then becomes manifest why we shall base on the HQEFT of QCD rather than the HQET.

 \section{Effects of Quark-antiquark coupling terms and Consistency of HQEFT}

The Lagrangian (\ref{lagr}) contains the fields of quark and
antiquark explicitly. It is then natural to ask what is the new
effects of HQEFT with the quark-antiquark coupling terms. For
that, it is better to consider the concrete physical processes.
But for many physical processes, the initial and final states
contain no antiquark (or quark), and therefore in these cases it
is convenient to deal with only the effective quark (or antiquark)
field variables. From this consideration it is useful to further
derive a Lagrangian represented only by quark  (or antiquark)
field variables but effectively taking the contributions of
quark-antiquark coupling terms into account.

From the Lagrangian presented in Eqs.(\ref{lagr})-(\ref{lzf}) it is easy to get
the relations between quark and antiquark field variables:
\newcommand{\LEFT}[1]{\overleftarrow{\hat{#1}}}
\begin{eqnarray}
&&(\CDP -m_Q )\QVHF + [i\DSC (\MQ-i\DSP)^{-1}(i\CDP -\MQ)]\QVHZ =0 , \\
&& \QVHFB (\CDPL - m_Q) + \QVHZB [i\DSCXL
    (\MQ-i\oDSP)^{-1} (i\CDPL -\MQ)] = 0 ,
\end{eqnarray}
or
\begin{eqnarray}
 \label{eq:new1a}
&& \QVHF=-(\CDP -m_Q )^{-1} [i\DSC (\MQ-i\DSP)^{-1}(i\CDP -\MQ)]\QVHZ ,\\
\label{eq:new1b}
 && \QVHFB=-\QVHZB [i\DSCXL (\MQ-i\oDSP)^{-1}
(i\CDPL -\MQ)] (\CDPL - m_Q)^{-1}.
\end{eqnarray}
Integrating out the heavy antiquark fields is equivalent to
substituting Eqs.(\ref{eq:new1a}), (\ref{eq:new1b}) into
Eqs.(\ref{lagr})-(\ref{lzf}). Consequently, we arrive at the
effective Lagrangian
\begin{eqnarray}
\label{eq:20}
{\cal L}^{(++)}_{eff}&=& {\cal L}^{(++)}_{Q,v} + \tilde{\cal L}^{(++)}_{Q,v},
\end{eqnarray}
which is nothing but representing the Lagrangian (\ref{lagr}) in
terms of only quark variables. The second part $\tilde{\cal
L}^{(++)}_{Q,v}$ comes from the contributions of integrating out
antiquark. Its explicit form is found to be
\begin{eqnarray}
\tilde{\cal L}^{(++)}_{Q,v} &=& \langle {\cal L}^{(--)}_{Q,v}+
    {\cal L}^{(+-)}_{Q,v}+{\cal L}^{(-+)}_{Q,v}\rangle \vert_{\hat{Q}^{(-)}_v \to \hat{Q}^{(+)}_v} \nonumber \\
&=&-\QVHZB
i\DSC(\MQ-i\DSP)^{-1}i\DSC(\MQ-i\DSP)^{-1}(i\CDP-\MQ)\QVHZ .
\end{eqnarray}

  After removing the large mass term in the above effective Lagrangian, we then obtain the following
  compact form
 \begin{equation}
 \label{complarg}
  {\cal L}^{(++)}_{eff} = \qpvb i\CDPN {1\over i\DSP} i\CDPN \qpv
 \end{equation}

  Note that up to now we do not yet make any approximation except making redefinitions for the quark fields and
integrating out the field component $R_v$ and antiquark field
$\hat{Q}^{(-)}_v$. So that the above effective Lagrangian should
still characterize the necessary features of QCD.

 To be manifest, we may reexpress the above Lagrangian in the following explicit form
 \begin{eqnarray}\label{newLeff}
{\cal L}^{(++)}_{eff} &=& \qpvb \Big\{i\DSP+ 2\cdot
\frac{1}{2\MQ}(i\DSC) \Big(1 - {i\DSP\over 2 \MQ}\Big)^{-1}
(i\DSC)\Big\}\qpv \nonumber
\\ &&+ \frac{1}{4\MQ^2}\qpvb(i\DSC) \Big(1 - {i\DSP\over 2
\MQ}\Big)^{-1} (i\DSC) \nonumber
\\
&&\times \frac{1}{i\DSP}(i\DSC) \Big(1 - {i\DSP\over 2 \MQ}\Big)^{-1} (i\DSC)\qpv .
\end{eqnarray}

 It is easily seen that the above effective Lagrangian for quark fields is distinguishable from the effective Lagrangian
 adopted in the HQET. There are actually two additional terms which arise from the contributions
 by integrating out
 the antiquark fields. One term is at $1/m_Q$ order, it exactly coincides with the one in ${\cal L}^{(++)}_{Q,v}$.
 This explains the factor of two in the second term of the curl bracket in
 the above effective Lagrangian. The other term is formally
 at $1/m_Q^2$ order but with the operator $\DSP$ in the denominator,
 which has to be treated carefully for
 different physical processes when making $1/m_Q$ expansion.
 Their physical consequences will be discussed
 in detail below. Such two additional terms explicitly represent new effects of HQEFT vs. HQET.

  Before proceeding, we shall first check the consistency of the above HQEFT.  Since the following issue naturally
  arises: as the above Lagrangian is exact without making any approximations, it should also recover the case for
  the free quark fields. The question is then how to understand the additional two terms
  formally at order $1/m_Q$ and $1/m_Q^2$.
  The answer is positive as one can see below.

   For the free quark case, there is no coupling between quark and gluon. One can simply take the
  coupling constant to be zero and make the replacement
  \beq
 \label{derivch}
&&D^\mu_{\|} \to \partial^\mu_{\|} = v^{\mu} v\cdot \partial , \nonumber \\
&&D^\mu\hspace{-0.3cm}_\bot \to \partial^\mu\hspace{-0.3cm}_{\bot} =
  \partial^{\mu} - v^{\mu} v\cdot \partial  , \nonumber \\
&&\CDPN \to i\partial\hspace{-0.2cm}\slash_v =
i\PSP + {1\over 2m_Q}i\PSC^2(1 + {i\PSP \over 2m_Q})^{-1}.
 \eeq
 Thus the Lagrangian for the free quark field is simply obtained from
Eqs.(\ref{complarg})-(\ref{derivch}),
\begin{eqnarray}
\label{lagfree} {\cal L}^{(++)}_{eff} &=& \qpvb
i\partial\hspace{-0.2cm}\slash_v {1\over i\PSP}
i\partial\hspace{-0.2cm}\slash_v \qpv\nonumber \\ &=& \qpvb
\Big\{i\PSP+ 2\cdot \frac{1}{2\MQ}(i\PSC)^2 \Big(1 + {i\PSP\over 2
\MQ} \Big)^{-1} \Big\} \qpv \nonumber \\ &&+
\frac{1}{4\MQ^2}\qpvb(i\PSC)^4 \Big(1 + {i\PSP\over 2 \MQ}
\Big)^{-2} \frac{1}{i\PSP}\qpv .
\end{eqnarray}
Notice the factor of two at the $1/m_Q$ order in the first term
and also the denominator $i\PSP$ in the second term which is
 formally at $1/m_Q^2$ order. It appears to be quite unusual. We
shall see that the second term is not truly at the order of
 $1/m_Q^2$ for the free quark field due to the denominator $i\PSP$.
Its leading contribution is actually
at $1/m_Q$ order with magnitude being half to the $1/m_Q$ order part in the first term but with opposite sign.

 To be convenient, let us turn to the momentum space, $i\partial\hspace{-0.25cm}\slash_v$ is represented as
\begin{eqnarray}
\label{DSV} v\cdot k + { k^2 - (v\cdot k)^2 \over 2m_Q}
  \Big(1 + {v\cdot k\over 2 m_Q} \Big)^{-1} .\nonumber
\end{eqnarray}
Note that when taking $v=(1, 0, 0, 0)$, it is simplified to be
\begin{eqnarray}
k^0 - {\mbox{\bf k}^2 \over 2 m_Q} \Big(1 + {k^0 \over 2m_Q} \Big)^{-1}. \nonumber
\end{eqnarray}
 As the free quark is on mass-shell, taking $v = (1, 0, 0, 0)$ for simplicity, one has
\begin{eqnarray}
\label{onshell1}
&&P^2 = (m_Q v + k)^2 = m^2_Q, \rightarrow k^0 - {\mbox{\bf k}^2 \over 2 m_Q}(1 + {k^0 \over 2m_Q})^{-1} = 0  \nonumber \\
\label{onshell2}
&&\mbox{or} \hspace{2cm} k_0 = \sqrt{m^2_Q + \mbox{\bf k}^2} - m_Q = {\mbox{\bf k}^2 \over 2 m_Q} +
O\Big({1\over m^2_Q}\Big).
\end{eqnarray}
In the coordinate space, Eq.(\ref{onshell1}) can be represented in the form of operators as
\begin{eqnarray}
\label{onshell3}
  i\PSP =- {1\over 2m_Q}i\PSC^2(1 + {i\PSP
\over 2m_Q})^{-1} = -{(i\PSC)^2 \over 2m_Q} + O\Big({1\over
m^2_Q}\Big).
\end{eqnarray}
Relation (\ref{onshell3}) means that when the quark is on
mass-shell, the longitudinal and transverse momenta are not at the
same order in the $1/m_Q$ expansion. In fact, the operator $i\PSP$
should be treated at the order of $1/m_Q$. More exactly it is at
the same order of $(i\PSC)^2/{2m_Q}$.

Substituting the relation (\ref{onshell3}) into the second term of the Lagrangian in Eq.(\ref{lagfree}), one has
\begin{eqnarray}
\label{freelagr} {\cal L}^{(++)}_{eff} &=& \qpvb
\Big\{i\PSP+\frac{1}{\MQ}(i\PSC)^2 \Big(1 + {i\PSP\over 2
\MQ}\Big)^{-1} \Big\}\qpv \nonumber \\ &-&
\frac{1}{4\MQ^2}\qpvb(i\PSC)^4 \Big(1 + {i\PSP\over 2
\MQ}\Big)^{-1}
  \Big[{(i\PSC)^2\over 2 \MQ}\Big]^{-1}\qpv   \nonumber \\
  &=& \qpvb \Big\{ i\PSP + {1\over 2m_Q}i\PSC^2(1 +
{i\PSP \over 2m_Q})^{-1} \Big\} \qpv .
\end{eqnarray}
It is seen that the resulting Lagrangian (\ref{freelagr}) is
self-consistent with the on mass-shell condition (\ref{onshell1}) or
(\ref{onshell3}). This becomes the known Lagrangian for a free
quark field. One then arrives at a consistent check for HQEFT in
the free quark case.

\section{HQEFT as a large component QCD of heavy quarks}

  We now apply the exact formulation of HQEFT to the heavy quark systems with $p \ll m_Q$.
  For this case, one may expand the HQEFT Lagrangian (\ref{newLeff}) into a series in powers
of the inverse heavy quark mass. Practically, in studying certain
physical processes it is reliable to consider only some lower
order contributions in the Lagrangian but neglect the higher order
$1/m_Q$ corrections. In order to correctly perform this expansion,
however, one must keep an eye on the physical conditions related
to the concrete processes. In particular, the longitudinal and
transverse momenta of the heavy quark may be at different powers
for different physical processes. Correspondingly, the operator
$\dsp$ and $\DSC$ may be treated as operators in different powers
in the expansion. To be more explicit, we consider two interesting
physical systems which are corresponding to two typical
conditions.

  We first consider the heavy quark within heavy quarkonia systems,
   which is known to be well treated by NRQCD. In fact, such a heavy quark
   is considered to be nearly on mass-shell. This implies the operator relation for the $1/m_Q$ expansion
   \begin{equation}
   i\DSP  = - \frac{(i\DSC)^2}{2\MQ}  + O\Big(\frac{1}{m_Q^2} \Big) .
   \end{equation}
  With this condition, it is not difficult to show that the effective Lagrangian (Eq.(3.8))
  has the following form in terms of the $1/m_Q$ expansion
 \begin{eqnarray}\label{newLeffNR}
{\cal L}^{(++)}_{eff} & = &
 \qpvb \Big\{i\DSP+\frac{1}{2\MQ}(i\DSC)^2 \Big\}\qpv  + O\Big(\frac{1}{m_Q^2}
  \Big).
\end{eqnarray}
Taking $v = (1, 0, 0, 0)$, one has
\begin{eqnarray}
{\cal L}^{(++)}_{eff}
 = \qpvb \Big\{i {\partial \over \partial t} + {{ \bf \nabla}^2 \over 2m_Q}
  \Big\}\qpv + O\Big({1\over m^2_Q} \Big).
\end{eqnarray}
 which does result in an NRQCD.

 We now consider another interesting case with heavy quark being slightly off mass-shell.
 This is considered to be the case for a single heavy quark within the heavy hadron.
 The magnitude of the off mass-shell is usually assumed to be at the order of bounding energy.
Namely it satisfies the off shell condition
\begin{eqnarray}
{P^2 - m^2_Q \over 2 m_Q} \sim \bar{\Lambda},\nonumber
\end{eqnarray}
Taking $P= m_Q v + k $ with $v= (1, 0, 0, 0)$, one has
\begin{eqnarray}
k^0 \sim \vert { \bf k}\vert \sim \bar{\Lambda} , \nonumber
\end{eqnarray}
 which has been considered in both HQET and HQEFT.
$\bar{\Lambda} \sim 500$ MeV is the typical bounding energy of
heavy hadrons. In general, one may take the corresponding
approximation in the operator form:
\begin{eqnarray}
\langle i\dsp \rangle \sim \langle i\DSC \rangle \sim \bar{\Lambda},
\end{eqnarray}
which implies that the operators corresponding to longitudinal and transverse momenta are taken to be at the same
order in the $1/m_Q$ expansion.

With the above considerations, the Lagrangian (\ref{newLeff}) can be expanded in terms of $1/m_Q$ into the following form
    \begin{eqnarray}
    \label{efflagr}
      {\cal L}^{(++)}_{eff}= {\cal L}^{(0)}_{eff}+{\cal L}^{(1/\MQ)}_{eff}
    \end{eqnarray}
with
    \begin{eqnarray}
    \label{efflagr0}
     {\cal L}^{(0)}_{eff} & = & \qpvb(i\DSP)\qpv ,  \\
    \label{efflagrm}
     {\cal L}^{(1/\MQ)}_{eff} & = & 2\ \frac{1}{2\MQ}\qpvb(i\DSC)^2\qpv +
   2\ \frac{1}{4\MQ^2}\qpvb i\DSC (i\DSP)i\DSC \qpv \nonumber \\
           && + \frac{1}{4\MQ^2}\qpvb (i\DSC)^2 \frac{1}{i\DSP}(i\DSC)^2 \qpv
           +O\Big(\frac{1}{\MQ^3} \Big) ,
    \end{eqnarray}
where ${\cal L}^{(0)}_{eff}$ is the leading term and possesses the
spin-flavor symmetry. The second part ${\cal L}^{(1/\MQ)}_{eff}$
contains the remaining terms which are the spin-flavor symmetry
breaking ones but suppressed by $1/m_{Q}$. Note that the present
case differs from the first case of the nearly on mass-shell heavy
quark. The third term in the rhs. of Eq.(\ref{efflagrm}) is now
regarded as at the order of $1/m^2_Q$. So the first term in the
rhs. of Eq.(\ref{efflagrm}) is now regarded as the total $1/m_Q$
order correction to the leading order Lagrangian (\ref{efflagr0}).
Obviously, the factor of two in this case can be understood
easily. It makes the HQEFT different from the HQET when applying
to the physical systems in which the single heavy quark within a
hadron is slightly off mass-shell, i.e., $\langle i\dsp \rangle
\sim \langle i\DSC \rangle \sim \bar{\Lambda}$. Their physical
effects will be discussed below.

  The above analysis shows that HQEFT provides a simple and systematic framework when the physical processes allow to
  make the expansions in terms of the inverse powers of heavy quark mass $m_Q$. In this sense,
  the HQEFT is nothing but a large component QCD of heavy quarks.

\section{ New features of HQEFT VS. HQET }

The quark-antiquark coupling terms in HQEFT should exhibit their
effects in its phenomenological applications to physical
processes. Some detailed analyses have been made in references
\cite{W1}-\cite{W10}. Here we revisit the most important features
of HQEFT with paying attention to the comparison with HQET.  In
general, HQEFT will lead to the results deviating from those in
HQET starting at $1/m_Q$ order. Especially, the HQEFT can largely
simplify the evaluations of hadronic matrix elements in comparison
with the HQET.

 First of all, a straightforward and automatic consequence in HQEFT of QCD is that at
zero recoil point the weak transition matrix elements between
ground state heavy mesons do not receive $1/m_Q$ order
corrections. Of course this was known to the so-called Luke's
theorem in HQET. However, in HQET one has to make a carefully
proof with the aid of the equation of motion $iv\cdot D Q^{(+)}_v
 = 0 $ to arrive at this statement. Also HQEFT of QCD is automatically
 reparametrization invariant\cite{YLW} as it is derived directly from QCD,
 whereas in HQET one needs to impose certain conditions in order to insure the
 reparametrization invariance\cite{RI1,RI2}.

 The HQEFT of QCD displays its advantage in the applications to the extraction
of the important CKM matrix element $|V_{cb}|$. It is known that
$|V_{cb}|$ can be extracted by studying $B\to D^* l\nu$ and $B\to
Dl\nu$ decays. In HQET the latter decay channel $B\to Dl\nu$
provides a less precise one in determining $|V_{cb}|$. Besides the
experimental reasons, an important reason for this is that the
Luke's theorem in HQET no longer suppresses the $1/m_Q$
corrections for the $B\to Dl\nu$ decay width. This may induce large
theoretical uncertainty to the extracted $|V_{cb}|$. The situation
becomes quite different in the HQEFT of QCD, where the suppression
of $1/m_Q$ corrections is valid also for the $B\to Dl\nu$ channel.
 As a consequence, HQEFT of QCD provides a more precise way than
HQET to extract $|V_{cb}|$.

It is noted that in HQEFT the $1/m_Q$ corrections to heavy meson
masses can be related to the $1/m_Q$ corrections to weak
transition matrix elements between the relevant heavy mesons.
While in HQET these are treated separately and characterized by
different sets of functions (factors).

 It is also very interesting to observe that heavy hadron dynamics is much simplified
in HQEFT than in HQET. Specifically, fewer independent functions
(factors) are needed in HQEFT than in HQET to characterize
the finite mass corrections. For instance, in HQET one needs 6
wave functions $\xi_i$ and $\chi_i$ $(i=1,2,3)$ to parametrize
the $1/m_Q$ order corrections to weak transition matrix elements
between heavy mesons, and needs other 2 factors $\lambda_1$ and
$\lambda_2$ for the $1/m_Q$ order corrections to meson masses. In
HQEFT the $1/m_Q$ corrections from current expansion and from
insertion of $1/m_Q$ order Lagrangian are attributed to the same
set of 3 wave functions $\kappa_i$ $(i=1,2,3)$; and the $1/m_Q$
order corrections to meson masses are also naturally related to
the zero recoil values of $\kappa_1$ and $\kappa_2$.
Instead of the 8 functions (factors) needed in HQET, one
thus needs only 3 independent functions in HQEFT.

In studying inclusive bottom hadron decays,
 the mass quantity entering into decay rates in the HQEFT formulation
 is no longer the heavy quark mass $m_Q$ but the ``dressed heavy quark"
mass $\hat{m}_Q=m_Q+\bar{\Lambda}$, which is related to the hadron mass at a high order $1/m_Q$ corrections, i.e.,
$\hat{m}_Q = m_H [1 + O(1/m_Q^2) ]$. In this way the inclusive decay rate formula of heavy hadrons is found to
have no $1/\hat{m}_Q$ order correction when it is given by the physical heavy hadron mass. This is because the
 HQEFT of QCD allows us to perform the heavy quark expansion at the point of the ``dressed heavy quark"
mass  $\hat{m}_Q$ instead of $m_Q$, which successfully suppresses the next to leading order contributions of the
expansion. Consequently, the HQEFT of QCD provides a precise determination of $|V_{cb}|$ and $|V_{ub}|$ in
 the inclusive bottom hadron decays.
 In particular, it can also present a well explanation for the puzzle of the bottom hadron life time
 differences, i,e., the ratios $\tau(B^0_s)/\tau(B^0)$ and $\tau(\Lambda_b)/\tau(B^0)$.

  In the above we briefly describe only some of the most important
features of HQEFT as compared to HQET. Actually, HQEFT of QCD has also been applied to study the issues such as
the charm counting in the bottom hadron systems, the heavy meson decay constants, heavy-to-light meson decays and so on. These applications have shown the power of HQEFT
for its consistency. All of them are attributed to the correct treatment for the quark-antiquark coupling terms.
 For more detailed discussions on these and other features, we refer to the refs.\cite{W1}-\cite{W10}.
  Note that the above interesting features of HQEFT cannot be obtained from the usual HQET
  through a simple redefinition for the heavy quark fields.

In conclusion, HQEFT is nothing but a large component QCD of heavy quarks. Its consistency and completeness become
manifest, which strongly suggests that we shall base on the HQEFT of QCD rather than the usual HQET to investigate
heavy quark systems beyond the leading order of $1/m_Q$ expansion.
We hope that the interesting features and consistent results
concerning the phenomenological applications of HQEFT will invoke
more and more attention to this effective theory.
\\
\\
\\

{\bf Acknowledgement}

 The authors (WYW and YLW) would like to thank  K. Hagiwara for stimulating discussions and conversation.
  They are also grateful to Y.Q. Chen, Y.B. Dai, C.S. Huang, T. Huang, X.Y. Li, Y.P. Kuang
  and many colleagues for useful discussions. This work was supported in part by the key projects
  of Chinese Academy of Sciences and National Science Foundation of China.

\end{document}